# DESIGN, INSTALLATIION AND INITIAL COMMISSIONING OF THE MTA BEAMLINE*

C. D. Moore[!], J. Anderson, F. Garcia, M. Gerardi, T. Kobilarcik, C. Johnstone, M. Kucera, M. Kufer, D. Newhart, I. Rakhno, and G. Vogel, FNAL, Batavia, IL 60510 U.S.A.

*Abstract*

A new experimental area designed to develop, test and verify muon ionization cooling apparatus using the 400-MeV Fermilab Linac proton beam has been fully installed and is presently being commissioned. Initially, this area was used for cryogenic tests of liquid-hydrogen absorbers for the MUCOOL R&D program and, now, for high-power beam tests of absorbers, high-gradient rf cavities in the presence of magnetic fields (including gas-filled cavities), and other prototype muon-cooling apparatus. The experimental scenarios being developed for muon facilities involve collection, capture, and cooling of large-emittance, high-intensity muon beams--~$10^{13}$ muons, so that conclusive tests of the apparatus require full Linac beam, which is $1.6 \times 10^{13}$ p/pulse. To support the muon cooling facility, this new primary beamline extracts and transports beam directly from the Linac to the test facility. The design concept for the MuCool facility is taken from an earlier proposal [1], but modifications were necessary to accommodate high-intensity beam, cryogenics, and the increased scale of the cooling experiments. Further, the line incorporates a specialized section and utilizes a different mode of operation to provide precision measurements of Linac beam parameters. This paper reports on the technical details of the MuCool beamline for both modes.

## INTRODUCTION

A new experimental area designed to develop, test and verify muon ionization cooling apparatus using the 400-MeV Fermilab Linac proton beam has been completed and is now being commissioned (Figure 1). This area is being used to test prototype muon cooling apparatus including liquid-hydrogen absorbers and high-power beam tests of rf cavities and superconducting solenoids for the MUCOOL R&D program. Since the experimental scenarios being developed for muon facilities involve large-emittance, high-intensity muon beams - ~$10^{13}$ muons, conclusive tests of the apparatus require up to full Linac beam intensity, or $1.6 \times 10^{13}$ p/pulse.

The new primary beamline extracts and transports beam directly from the Linac to the test facility. The beamline was designed with a dispersion-free straight insertion and supports a dual mode of operation. The two modes, termed emittance and experiment, provide, respectively, an accurate measurement of Linac beam properties and customized beam to the experiments. The following sections detail beamline optics, operation and performance in the two modes.

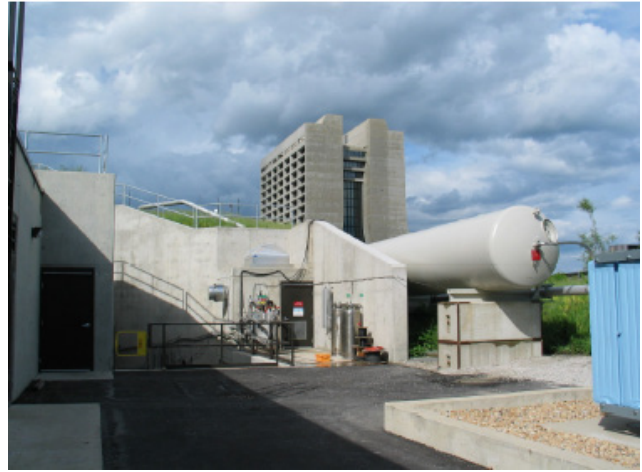

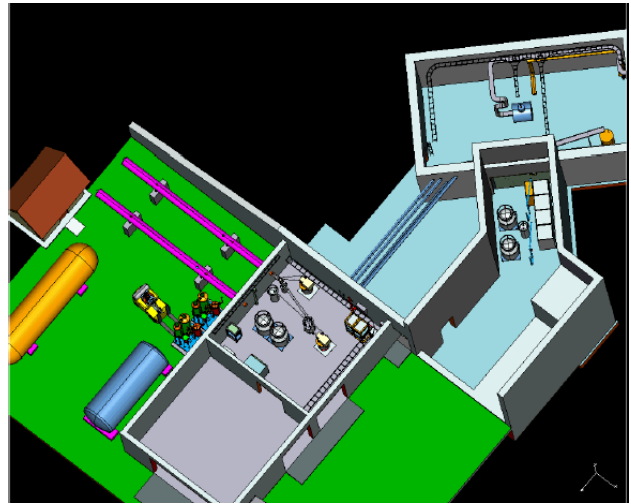

Figure 1. Exterior photograph and drawing of the civil construction for the MTA facility.

## BEAMLINE DESIGN

The MuCool beamline must operate parasitically to the Fermilab HEP program. Beam is therefore fully extracted on a single 15 Hz tick which corresponds to the maximum duty cycle of the Fermilab Linac. Intensity in the MTA beamline is controlled by changing the repetition rate (up to 15 Hz) of a fast extraction C magnet in combination with an electrostatic beam chopper, which can vary the Linac pulse length between 20 and 50 μsec. This corresponds to a pulse intensity of $0.64 - 1.6 \times 10^{13}$ protons. The minimum pulse length is determined by stable performance of the rf feedback systems in the Linac.

___________________________________________
Work supported by the Fermilab Research Alliance, under contract DE-AC02-07CH11359 with the U.S. Dept of Energy.
[!]cmoore@fnal.gov

In the emittance mode, beam is delivered to an intermediate beam absorber at a maximum rate of 10 pulses/minute. In the beamline design, the primary-beam enclosure shielding (12') was advantageously used to implement a long, 10m straight reserved for a magnet-free diagnostic section to measure transverse Linac beam properties (Figure 2, top). Three multiwire profile monitors are installed at the upstream, center, and downstream with 1 mm pitches in the outer two wires and 0.5 mm in the center wire to provide accurate, detailed profiles. Quadrupole-triplet telescopes on either side of the straight form an intermediate waist and further allow variable phase advance across the straight thus providing a flexible and powerful basis for beam tomography. The set of optics for emittance measurements is shown in Figure 2 (bottom) for the entire beamline (the normalized Linac beam emittance is about $10\pi$ mm-mr and the nominal vacuum pipe aperture is 3.25"). Note the long magnet-free straight which extends from about 31 to 41 m enables a virtually systematic-free measurement of Linac beam properties, in particular emittance.

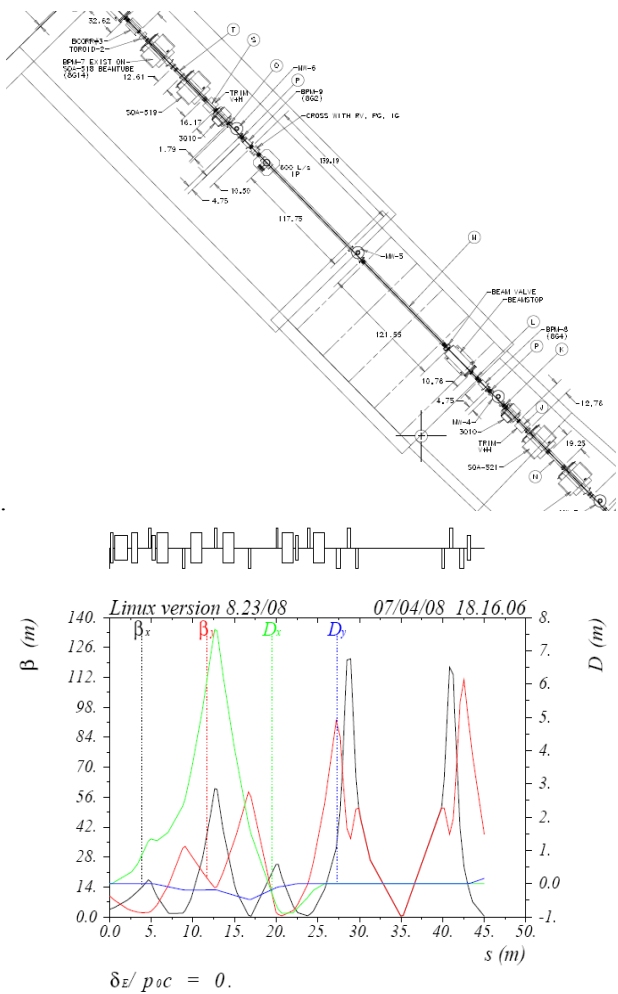

Figure 2. The long straight section instrumented for beam tomography (top) and the emittance mode lattice (bottom).

Beam is delivered to the muon cooling apparatus in the experiment mode at an order of magnitude lower rate than in the emittance mode due to penetration and radiation shielding constraints related to the experimental hall. Several categories of experiments are supported on the experiment mode: complete beam interception (rf cavities with thick windows, for example), minimal interception with most beam transmitted to the final, high-intensity beam absorber (such as hydrogen absorbers or rf cavities with thin windows), or entire transmission of the beam to the final beam absorber. In the latter two cases, beam must be transported to the high-intensity beam absorber buried in the berm downstream of the experimental hall. The range in beam parameters that is available to experiments is detailed in Table 1. Further reductions in beam intensity up to an order of magnitude may be achieved through insertion of pinhole collimators in the experimental hall.

Table 1. Phase I beam parameters for the MTA facility

| Beam Specifications | Min | Max |
|---|---|---|
| Beam Size (±3σ, or full width) | 1 cm | 5 cm |
| Beam Divergence (±3σ, full width) | 2 mr | 0.4 mr |
| Number of Pulses per Second | - | 60/hr ≤15Hz |
| Number of Proton/pulse x $10^{13}$ | 0.64 | 1.6 |
| Pulse Duration | 20 μs | 50 μs |

## OPERATION

The dual mode of operation is implemented operationally by a rate limiting device. To limit the hourly beam intensity to the MTA facility, a Power Supply Repetition Rate Monitor Interlock has been designed and installed. The repetition rate monitor interlock allows for the number of pulses and time interval to be specified, thus limiting the facility to "n" beam pulses over the "t" time interval.

The repetition rate monitor interlock system has been implemented with a Siemens Failsafe Programmable Logic Controller (PLC) that monitors two different signals from the extraction C-magnet power supply. The power supply SCR Trigger Pulse used to initiate the extraction magnet current pulse and I>50 Amps which directly measures the magnet current pulse. If the repetition rate monitor sees an excess of pulses over the specified time period from either of the input signals, the interlock will remove its redundant output permits to the MTA Critical Device Controller thereby disabling beam to the facility.

Since operation of the MuCool beamline involves direct control over the 400-MeV Linac beam, integration into the overarching accelerator control and protection system was required. A unique event was assigned to enable beam under the control of the master accelerator clock (TimeLine Generator). Although a complete integration has not been completed to date, a Linac event coupled to a permit system and operational safeguards is used to operate the beamline. Given the limited number of event slots (4) which characterize the present Linac control system, the Linac study pulse slot was selected and engineered to perform double duty for either Linac studies or for operation of the MuCool beamline. A master switch (keyed) selects the nature of this Linac event: either a Linac study pulse is enabled or beam is transported to the MuCool beamline. Eventually, full independent integration of MuCool-specific beam is planned.

## FIRST BEAM

First beam was established through the MuCool beamline in fall, 2008. Although many of the MuCool beam diagnostics were still under development, 100% beam extraction was confirmed by an upstream toroid after adjustment of the C-magnet strength. Further, beam was fully extinguished in the other "straight-ahead" lines derived from the Linac. Beam was successfully transported to the beamstop at the upstream side of the shield wall.

However, significant cross talk between the 400-MeV Booster injection line and the MuCool beamline remained. In addition to mu-metal shielding, a permanent magnet has since been installed replacing an electromagnet. Preliminary commissioning has just started with the permanent magnet. Independent and parasitic operation of the MuCool beamline has been confirmed with the permanent magnet in place.

## RADIATION ASSESSMENTS

The complex shielding assessment is almost complete entailing the emittance mode and three categories of the experiment mode. Full commissioning will take place after the July, 2010 accelerator maintenance shutdown. Given the complex nature of the facility extensive radiation simulations were required and include using the code MARS[2].

Normally, the total number of protons delivered is dictated by the physics of an experiment, and since experiments conventionally use orders of magnitude less intensity and highly-irradiated components—such as beamline elements and targets—are rarely worked on or modified, hands-on restrictions are not generally a defining issue for an experiment, but rather a shielding one. Generally, there is low residual activation of experimental detectors, and, once an experiment is up and running, accesses, work, and occupation of the hall is automatically limited.

This is not the case for the high-intensity facility proposed for MuCool R&D. With a high-intensity primary beam, residual activation of the components has become a defining issue for the experimental program. Figure 3 represents a MARS simulation of Linac full-intensity pulses stopped in an experimental device that represents a 100% interaction-length target (a high-pressure, gas-filled rf cavity with thick windows, for example). At 1 pulse/minute significant residual radiation is generated even for 1 hour of beamtime.

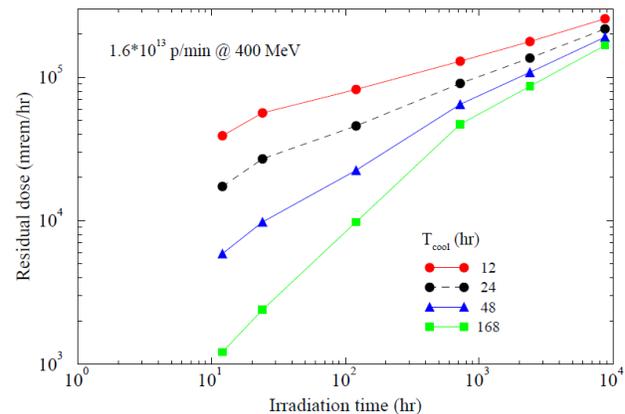

Figure 3. Residual activation of a 100% interaction length experimental device for varying irradiation and cool-down periods.

Further, the occupancy of the hall is expected to be frequent as modifications naturally follow the development of the experimental apparatus, with maintenance on and change-out of irradiated components potentially routine. The guidelines and administrative control over radiation and dose levels, then, is very similar to those established for accelerator operation and maintenance, but with the added complication of high occupancy.

## SUMMARY AND PROSPECTS

The MuCool beamline installation is complete and first beam has been transported to the upstream face of the primary enclosure shielding. The shielding assessment is nearing completion and full commissioning to the first experiment, a gas-filled rf cavity, in the experimental hall is anticipated after the July, 2010 accelerator maintenance shutdown.

## REFERENCES

[1] FERMILAB-PUB-95-078 (Mar 1995).
[2] Fermilab TM-2248, May 2004 and TM-2305-AD.